# Designing for human–AI complementarity in K-12 education

Kenneth Holstein and Vincent Aleven

**Abstract:**
Recent work has explored how complementary strengths of humans and artificial intelligence (AI) systems might be productively combined. However, successful forms of human–AI partnership have rarely been demonstrated in real-world settings. We present the iterative design and evaluation of Lumilo, smart glasses that help teachers help their students in AI-supported classrooms by presenting real-time analytics about students' learning, metacognition, and behavior. Results from a field study conducted in K-12 classrooms indicate that students learn more when teachers and AI tutors work together during class. We discuss implications of this research for the design of human–AI partnerships. We argue for more participatory approaches to research and design in this area, in which practitioners and other stakeholders are deeply, meaningfully involved throughout the process. Furthermore, we advocate for theory-building and for principled approaches to the study of human–AI decision-making in real-world contexts.

Artificial intelligence (AI) systems are increasingly used to support human work in deeply social contexts such as education, healthcare, social work, and criminal justice. In these contexts, AI can automate routine parts of practitioners' work, while freeing up their time for activities they find more meaningful (*17, 28, 39*). AI can also help to scale up the delivery of social services and help humans make more informed and equitable decisions (*5, 16, 28*). Despite these benefits, modern AI systems are fallible and imperfect. If not carefully designed, AI risks rigidly scaling practices without sensitivity to local context, propagating harmful inequities, or automating away valuable human–human interactions (*1, 4, 8, 18*). To ensure that these systems do more good than harm, it is critical they are designed to bring out the best of human ability while also helping to overcome human limitations.

A rich line of research has explored the design of effective *human–AI partnerships*: configurations of humans and AI systems that draw upon the strengths of each (*6, 16, 19, 22, 36*). Such integrations of human and machine intelligence have sometimes been shown to be more effective than AI or humans working alone (*4, 16, 20, 28*). For example, successful partnerships have been demonstrated in radiology, where human radiologists and AI systems working collaboratively exhibited higher diagnostic performance than either in isolation (*28*). By contrast, in many studies, human–AI collaboration has failed to improve or even *harmed* task performance (*8, 29, 35*). For instance, Poursabzi-Sangdeh et al. (*29*), found that increasing human visibility into the way a machine learning model makes predictions had the effect of *decreasing* rather than increasing humans' ability to detect and correct for model errors, apparently due to cognitive overload. Similarly, Green & Chen (*8*) found that including crowdworkers in the loop in a criminal risk assessment task led to worse predictive performance than a model operating alone.

So, why do some human–AI partnerships succeed, while others fail? Oftentimes, partnerships fail due to a lack of human-centered design – for example, where humans are unable to usefully interpret what the AI system is telling them or are overwhelmed by the manner in which the information is presented (*29, 35*). In other cases, partnerships may fail due to ineffective pairings, where there is simply no reason to expect, upfront, that humans and AI systems will have

Human–Computer Interaction Institute, Carnegie Mellon University, Pittsburgh, PA 15213, USA.

complementary strengths to build upon. For example, many null or negative results have come from studies on Amazon's Mechanical Turk, where crowdworkers assist an AI on tasks that truly require expert-level domain knowledge, which we cannot expect an average crowdworker to have (*4, 24*). Finally, as demonstrated in recent work, human–AI partnerships may sometimes *appear* to fail due to inappropriate evaluations. For example, Buçinca et al. (*3*) observed that empirical studies of human–AI partnerships rarely evaluate performance on actual decision-making tasks. Yet commonly used evaluation criteria, such as measuring humans' ability to predict AI decisions in particular instances, provide limited insight into human–AI performance on authentic decision-making tasks.

This article presents a case study of an effective human–AI partnership, achieved through human-centered and participatory design, in a challenging context: K-12 education. Successful human–AI partnerships have rarely been demonstrated in real-world social settings. As an example of a domain in which human care for other human beings is central, education represents both a challenging domain and fertile ground for human–AI synergy. Throughout this case study, we illustrate three recommendations for the design of effective human–AI partnerships, which we expect will generalize to similar professional contexts such as healthcare, social work, and criminal justice. These include **(1)** taking a *participatory* approach to research and design, deeply involving practitioners in framing the problems to be addressed and in designing how a partnership will function; **(2)** iteratively measuring and shaping *human–AI decision-making* in real-world contexts; and **(3)** working towards a *theory of complementarity*: an understanding of what complementary strengths humans and AI systems hold in a given context, which can be used to guide the design of systems that combine these strengths.

## CASE STUDY

### Background

Our case study focuses on AI-supported K-12 classrooms, a context in which human teachers and AI systems already work side by side, although not typically in carefully designed partnerships. AI-based tutoring systems have a long history in interdisciplinary research and are increasingly being used in K-12 classrooms (*5*). As students work on complex problem-solving and other learning activities, these systems use AI plan recognition algorithms to respond to individuals' problem-solving strategies, solution paths, and errors. They adapt instruction to individual students' needs, based on real-time models (often machine-learned) that track students' behavior, their knowledge growth, their metacognitive and self-regulated learning abilities, and even their affective states.

Several meta-analyses have shown that AI tutors can help students learn more effectively than other forms of instruction, across a wide range of domains (*5*). However, the role teachers play in K-12 classrooms using AI-based tutoring software remains under-studied (*13, 21, 27*). Prior field studies have found that as students work with the software, teachers, circulating through the classroom, are freed up to provide one-on-one guidance to students in need of additional assistance, e.g., (*13–14, 21, 27, 33*). While these studies give us reason to suspect that teachers play important roles in mediating students' learning with AI-based tutoring software, our scientific understanding of how this mediation plays out in practice is very incomplete. The same can be said for our understanding of how AI systems might be designed to work with teachers more effectively, to support even greater student learning outcomes (*14, 28, 38)*.

The current case study describes the design and field evaluation of a more effective form of human–AI partnership for K-12 classrooms that use AI tutors. While prior research has explored the design of tools to support teachers during ongoing classroom instruction – such as learning analytics dashboards and classroom management software (*2, 33*) – this work has rarely targeted contexts in which teachers work alongside AI-provided instruction. Yet AI-supported classrooms raise unique challenges for teachers, and in turn, for the design of teacher support. For instance, AI tutoring software often personalizes the content and pacing of educational activities based on automated inferences about individual students' needs (*32*), which can in turn make it challenging for teachers to keep track of individual students' activities. Furthermore, given that AI tutoring software does not typically coordinate with teachers about how to sequence and pace students' trajectories through the curriculum, conflicts can arise between AI decision-making and a teacher's plans and objectives for the class (*14, 18, 32*).

In line with the first of our recommendations for the design of effective human–AI partnerships (*take a participatory approach to research and design*), a key goal of the current project was to actively involve teachers throughout all phases of the design of a new real-time support tool (*17*). Recent reviews of the literature on teacher support tools have noted that the design of these tools often appears to be guided more by the availability of existing technical solutions (*27, 33*) than by an analysis of what would help teachers the most. However, tools resulting from this approach often present information that teachers find difficult to productively act upon (*14, 33*). Thus, we wanted to begin the current project with a thorough exploration of teachers' information needs, to guide design.

**Design and development of Lumilo**
The initial, exploratory phases of the current project spanned a wide range of human-centered and participatory design activities, conducted with K-12 teachers who had previously used AI tutors in their classrooms (*17–18*). These activities included field observations in K-12 classrooms, directed storytelling exercises to understand teachers' past experiences using AI in the classroom, generative card sorting exercises to better understand challenges teachers face during AI-supported class sessions, and speed dating studies to explore multiple potential futures for the role of AI in education. As an example, to find out what information teachers wanted to have about their students in real-time during a class session, unconstrained by their notions of what is technologically feasible, we asked them what "superpowers" they would want (*14, 17*). We found that in teachers' current practice, much of the rich information they take in during AI-supported class sessions comes from "reading the classroom:" actively looking at their students' body language and facial expressions. As such, they emphasized that an effective tool would need to allow them to keep their eyes and ears on the classroom, augmenting rather than distracting from signals already used in their day-to-day practice (*2, 12, 14*).

Building upon the rich findings from these formative research activities, we next conducted an iterative series of design and prototyping studies with teachers. To engage teachers in the co-design process as these prototypes achieved greater technical fidelity and complexity (e.g., using authentic data and machine learned models), it became necessary to innovate on design and prototyping methods (*11–12, 15, 17–18*). For example, we developed a new prototyping method called Replay Enactments (REs). REs embed participants in immersive simulations based on actual data collected from field contexts, to make the consequences of algorithm design decisions more tangible to stakeholders, even if they know very little about AI. During a session, a member of the research team makes live changes to algorithmic elements of a systems' design based on stakeholder feedback (e.g., parameters of a machine-learned model), so they experience the consequences of their requested changes (*11–12, 17–18*). REs can reveal critical issues that conventional prototyping methods cannot surface (e.g., helping design teams observe the interplay between human and AI systems' dynamic decisions and errors). In our project, we ran simulations with teachers in classrooms or computer labs, without students, but with student work replayed in real time on the computer screens, while the teacher, informed by the tool prototype, would act out what they would do if this were a real classroom period. By doing so, we gave teachers a space to iterate on Lumilo's design, without risking harm to actual students in the process.

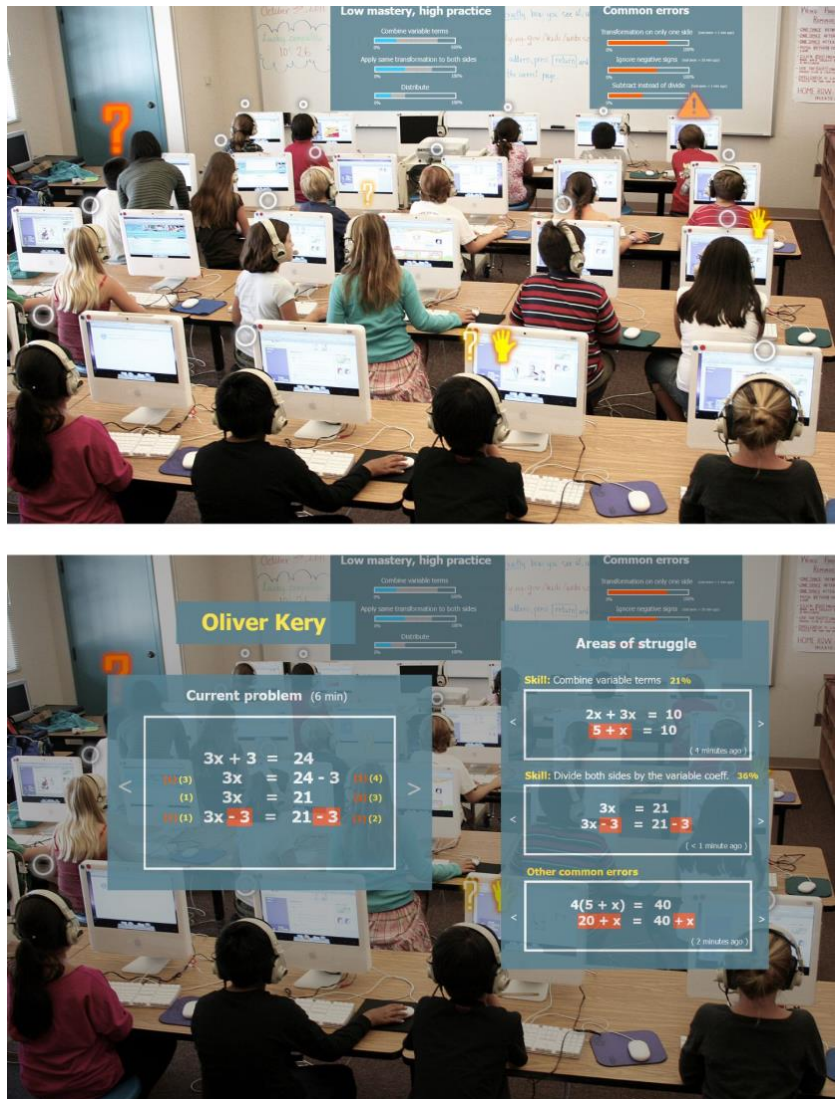

**Fig. 1.** Design mock-ups based on findings from low- to mid-fidelity prototyping sessions; from (*12*). (A) Teacher's default view of the class. Each student has an indicator display floating above their head, and class-level analytics displays are positioned at the front of the class. (B) "Deep-dive" screens shown if a teacher 'clicks' on an indicator. Note: student names shown in this figure are fabricated.

The prototype that emerged from this iterative design process was a pair of mixed reality smart glasses called Lumilo, which augment teachers' perceptions of student learning, metacognition, and behavior during AI-supported class sessions. When teachers glance across the room while wearing Lumilo, they can see mixed reality icons floating above each individual student's head (see Figure 1, A). These icons update throughout a class session based on real-time AI models embedded in the tutoring software, alerting teachers to situations in need of their attention. For example, if a student appears unlikely to master certain skills without additional help beyond the software, a red question mark icon would appear over the student's head. With such situations prioritized for teachers, they can make more informed decisions about *whom* to help and *when*.

The use of a wearable, heads-up display – mixed reality smartglasses, implemented using the Microsoft HoloLens – allowed teachers to keep their heads up and their attention focused on the classroom, rather than buried in a screen (*2, 12–13*). In addition to providing information at a glance, Lumilo can also display more detailed information about individual students upon a teacher's request. For example, if the AI tutor diagnoses that a student is struggling with particular skills at a given moment during class, Lumilo would display these diagnoses, together with concrete examples of recent errors the student had made on each skill (see Figure 1, B). Displaying these concrete, "raw" examples alongside the AI system's diagnoses proved to be very important to teachers, who would often use the examples to second guess the system's judgments and try to infer deeper underlying causes of student difficulties – an example of one form of complementarity between the AI and the teacher (*2, 17*).

**Iterative piloting: Measuring and shaping human–AI decision-making**

Prior to running a large-scale evaluation study with Lumilo in K-12 classrooms, we wanted to ensure that this form of human–AI partnership was likely to have a positive impact on students' learning. As discussed above, the initial designs of Lumilo were largely designed based on *teachers' beliefs* about the classroom situations that most required their attention. However, it is possible that teachers' intuitions are limited in this regard (*15*). Thus, to complement our co-design process, we ran an iterative series of pilot studies in both replayed classrooms (using REs) and live K-12 classrooms. In line with our second recommendation for the design of effective human–AI partnerships (*measure and shape human–AI decision-making in context*), we measured the impacts of particular designs on teachers' decision-making, iteratively refining Lumilo's design with the goal of guiding teacher decision-making in positive directions.

We developed and used Causal Alignment Analysis (CAA), a systematic approach to support the data-driven, outcome-oriented design of teacher–AI systems (*15*). The CAA approach asks technology designers to begin by specifying the educational goals and outcomes they wish to achieve (e.g., improving on particular measures of student learning or engagement), and then to work backwards by sketching out one or more hypothesized causal paths by which those outcomes might be achieved. For example, for a given outcome, a designer might specify (1) hypotheses regarding possible changes in student behavior that could support that outcome, followed by (2) possible changes in teacher behavior that might support the corresponding changes in student behavior, and finally (3) possible ways a teacher-facing AI tool might foster these changes in teacher behavior. These hypothesized causal paths may initially be informed by existing theory and empirical data, where available. When technology prototypes are tested in real-world contexts, data collected from these studies offer an opportunity not only to iterate on the design of the technology itself, but also to question and iterate upon the designer's hypothesized causal paths. By applying CAA over successive iterations, designers can iteratively refine their designs towards achieving their outcomes of interest.

Applying CAA in the iterative design of Lumilo meant first making our hypotheses, as researchers and designers, about Lumilo's mechanisms of action *explicit*, and then iteratively prototyping Lumilo in K-12 classrooms. During classroom pilots, we tracked teachers' activities, including how they allocated their time between different students throughout each class session. Using this data, we analyzed whether the tool was having desirable effects with respect to our hypothesized mechanisms of action, while simultaneously evaluating the plausibility of these hypotheses; see (*15*) for details of this analysis. Using CAA, we iteratively refined the design of Lumilo over a sequence of four pilot studies, with a total of 14 teachers, 15 classrooms, and 304 students. In the end, the resulting version of Lumilo appeared to direct teachers' attention where it was most needed in the classroom, as judged by classroom observation and causal modeling of how students learn with AI tutors (*9, 15*).

In line with our third recommendation for the design of effective human–AI partnerships (*work towards a theory of complementarity*), these pilot studies not only enabled design refinement based on quantitative metrics of teacher behavior (*15*), but also enabled rich qualitative observations that grew our understanding of how this human–AI partnership would play out in the real world. In turn, these field observations enabled us to better understand *why* the particular form of teacher–AI partnership facilitated by Lumilo might have a positive impact in the classroom (*9*).

Major themes that emerged from classroom observation were that the glasses helpfully alerted teachers to situations where their attention could be beneficial, and that teachers did indeed make in-the-moment decisions based on complementary data sources. For example, one teacher commented that, without the glasses, *"I wouldn't have known this student was doing this at this time."* In the moment, teachers would combine what they saw with their own eyes and ears with what the AI system was telling them about their students. As one teacher said, *"I would also use their body language to judge the situation, but the initial [alert] would help, so I know to go over there."* This use of complementary data often played out in interesting ways. For example, in one particularly memorable case, Lumilo alerted the teacher that a particular 7th grader may be off task in the software. However, based on what the teacher knew about this student, they perceived that this behavior was out of character. Therefore, rather than taking the alert at face value, the teacher initiated a conversation with the student, asking how the student was *feeling* that day. The student revealed that their significant other had broken up with them the weekend before. The teacher, in turn, gave the student permission to "take the day off" from math, if they wished (*9, 17*).

These examples illustrate a form of human–AI complementarity. The AI diagnosed a particular student behavior and alerted the teacher. Based on this information, the teacher then made a rich inference about the latent, underlying *cause* of the behavior, and responded with support and flexibility that an AI tutor could not provide (*2, 17*). More broadly, we observed that teachers often appeared to be very effective in helping students escape unproductive ruts in the AI software, with very brief, minimal guidance. As teachers circulated throughout their classrooms while using Lumilo, they spent an average of about 24 seconds per visit with each student, although teachers often visited the same student multiple times during a class; see (*15, 17*). In many of the cases we observed, rather than providing coaching on the math content itself, teachers complemented the AI tutors' instruction by motivating them to reflect on their activities (e.g., *"So what should you do next?"* or *"Why did you subtract x from the right side?"*), or by providing words of encouragement (e.g., *"I think you got this, you can do [the rest] on your own"*).

**In-the-wild evaluation study**

As the design iterations converged, we next conducted a study to better understand Lumilo's impacts on teacher behavior and student learning in AI-supported classrooms. We investigated the hypotheses that a teacher's use of Lumilo would enhance student learning in AI-supported classrooms, compared to helping students (a) without a teacher support tool ("business as usual") and (b) with mixed-reality glasses that provide only weak classroom monitoring support, without analytics. The latter condition made it possible to gauge any motivational or novelty effects that teacher monitoring might have on student learning, as observed in prior empirical work *(13, 35)*, so as to isolate the influence of Lumilo's analytics.

Participants were 343 middle school students, across 18 classrooms, 8 teachers, and 4 schools and school districts; for participant demographics, see (*17*). All participating teachers had previous experience using AI tutors in their classrooms and had at least 5 years of experience teaching math at a middle school level. Classrooms were randomly assigned to one of 3 conditions, stratifying within-teacher. The *No Glasses* condition represented "business as usual" for an AI-supported classroom. Teachers circulated throughout the classroom, peeking over students' shoulders without a teacher support tool. The *Glasses* condition provided a minimal form of classroom monitoring support: the teacher wore a stripped-down version of Lumilo, which did not show any analytics. However, teachers were still able to select individual students using their glasses to peek, at a distance, at what was currently displayed on that student's screen. Finally, in the *Glasses + Analytics* condition, the teacher used the full version of Lumilo. This version offered the remote screen monitoring functionality, at-a-glance visual indicators for each student based on real-time analytics, and detail screens indicating where students are struggling, as described above.

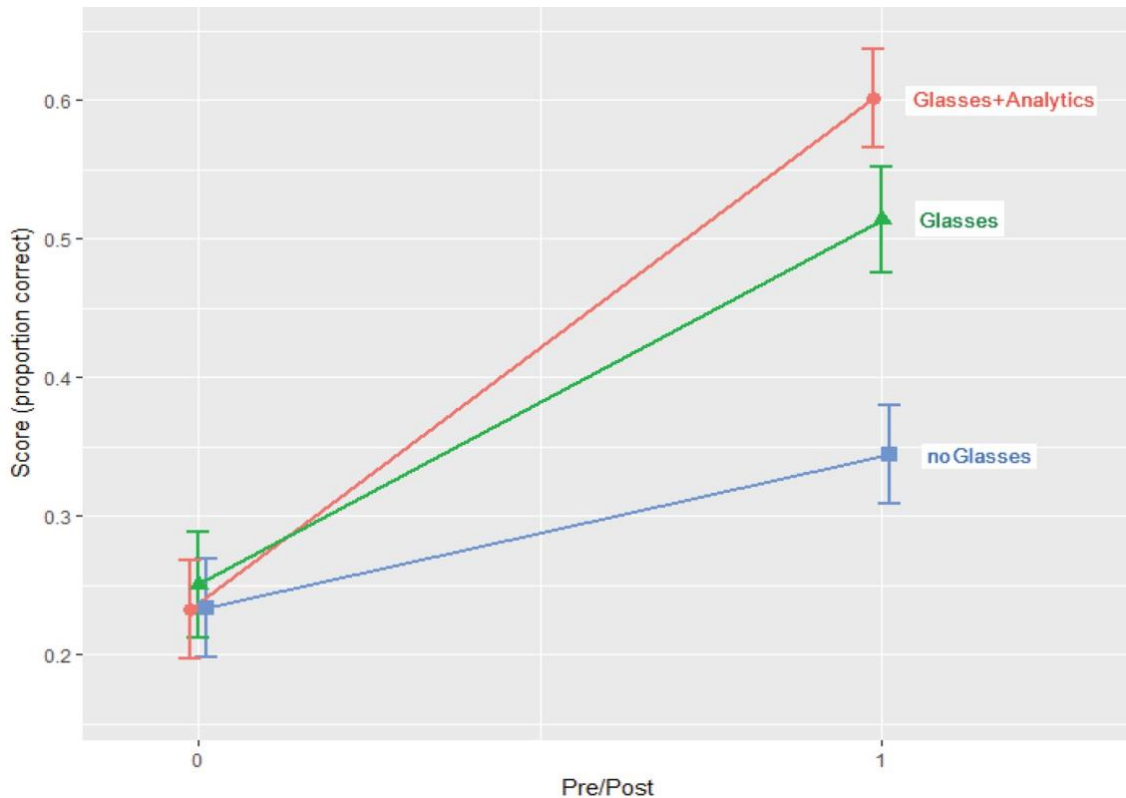

**Fig. 2.** Student pre/post learning gains, by experimental conditions ("Glasses + Analytics": Teacher uses Lumilo; from (*16*). "Glasses": Teacher wears reduced version of Lumilo, without analytics; "noGlasses": Teacher does not wear glasses at all). Error bars indicate standard error.

The study procedure was the same for all three conditions. Students took a 15-minute pre-test on linear equation solving. Students then worked with Lynnette, an AI tutor for linear equation solving, for two class sessions, while their teacher monitored the class and helped students. Finally, students took a 15-minute post-test. In addition to students' pre- and post-test scores, we tracked process data from individual students' interactions with Lynnette. We also used Lumilo to record a teacher's physical position in the classroom, relative to each student, moment-by-moment via the HoloLens' built-in sensors; see (*15–17*). Fifty-seven students were absent for one or more days of the study and were excluded from further analyses. Data were analyzed for the remaining 286 students, using hierarchical linear modeling; see (*16*) for details.

Analysis of the pre- and post-tests supported both of our hypotheses. First, a teacher's use of Lumilo enhances student learning, compared with business-as-usual for AI-supported classrooms, a high-bar control condition considering prior positive results regarding the effectiveness of AI tutoring software (*5*). Second, a teacher's use of real-time analytics had a positive effect on student learning, *above and beyond* any effects of the minimal classroom monitoring support provided in the *Glasses* condition. Thus, part of Lumilo's benefit was due to its real-time analytics, and part of it was due to the mere use of the glasses, even without any advanced analytics (see Figure 2). This was the first experimental study in the literature to demonstrate that a teacher–AI partnership, facilitated by real-time analytics from AI tutors, can enhance student learning outcomes.

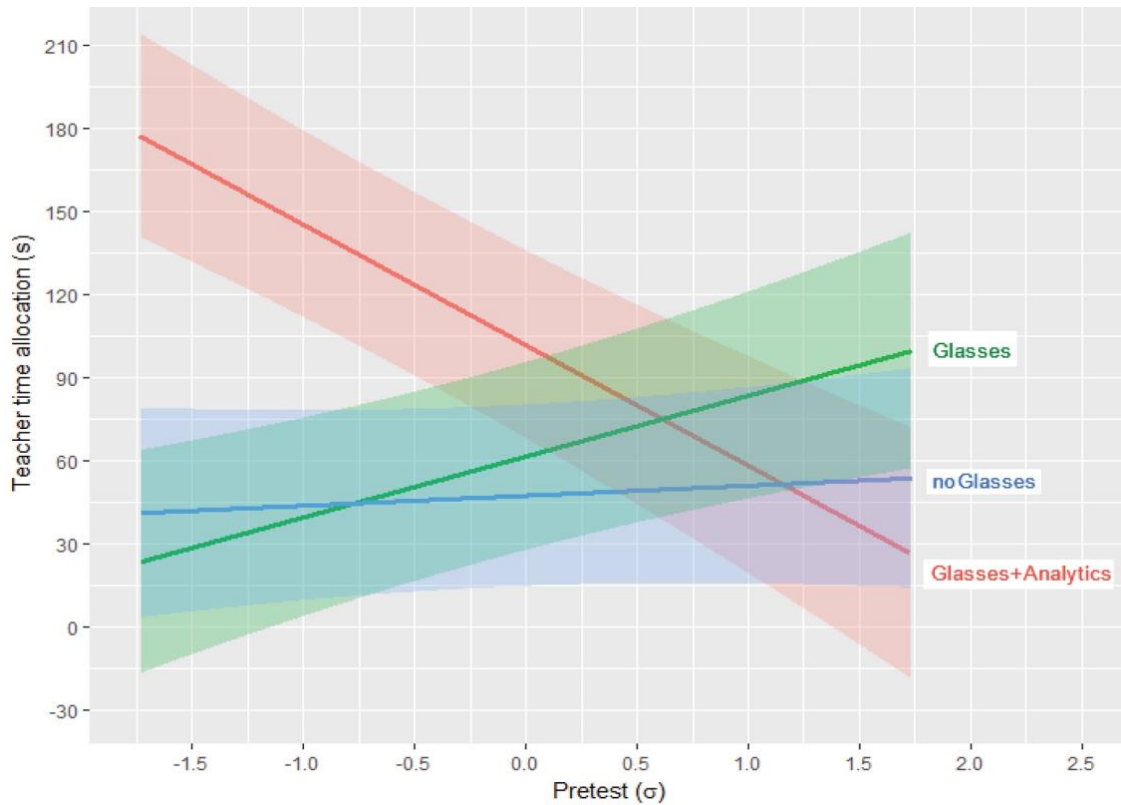

**Fig. 3.** Teacher attention allocation (in seconds) plotted by pretest scores, across 286 students. Lines indicate model mean values for each experimental condition; shaded regions indicate standard error; overlapping shaded regions indicate overlapping standard errors.

To better understand the mechanisms by which this effect may have arisen, we examined how teachers' allocation of time, across students of varying incoming knowledge, was influenced by experimental condition. We found that, compared with the *Glasses* and *No Glasses* conditions, teachers in the *Glasses + Analytics* condition tended to spend much more of their time working with students coming in with lower initial knowledge (as measured by the pre-test); see Figure 3. In turn, students with lower pre-test scores enjoyed greater growth in the Lumilo condition than in the other conditions, whereas those with high pre-test scores were not affected. In both the *Glasses* and *No Glasses* conditions, we observed the "rich get richer" trends that are sometimes observed in AI-supported classrooms and in education more generally: students coming in with higher initial knowledge tend to benefit more from working with the tutor (*30*). However, in the *Glasses + Analytics* condition, teachers' use of Lumilo attenuated these trends, reducing knowledge differences between students at post-test (see Figure 4). The notion that redirecting teachers' attention during personalized class sessions might benefit students' learning has found correlational support in prior studies (e.g., *13, 15, 26, 35*), yet had not received experimental support until the current study. Considering that the overall time a teacher can spend per student during these class sessions is still quite small, these results suggest that a little can go a long way, in terms of individualized teacher attention, especially when timely and targeted with the help of AI. For a more detailed report on these analyses, see (*16*).

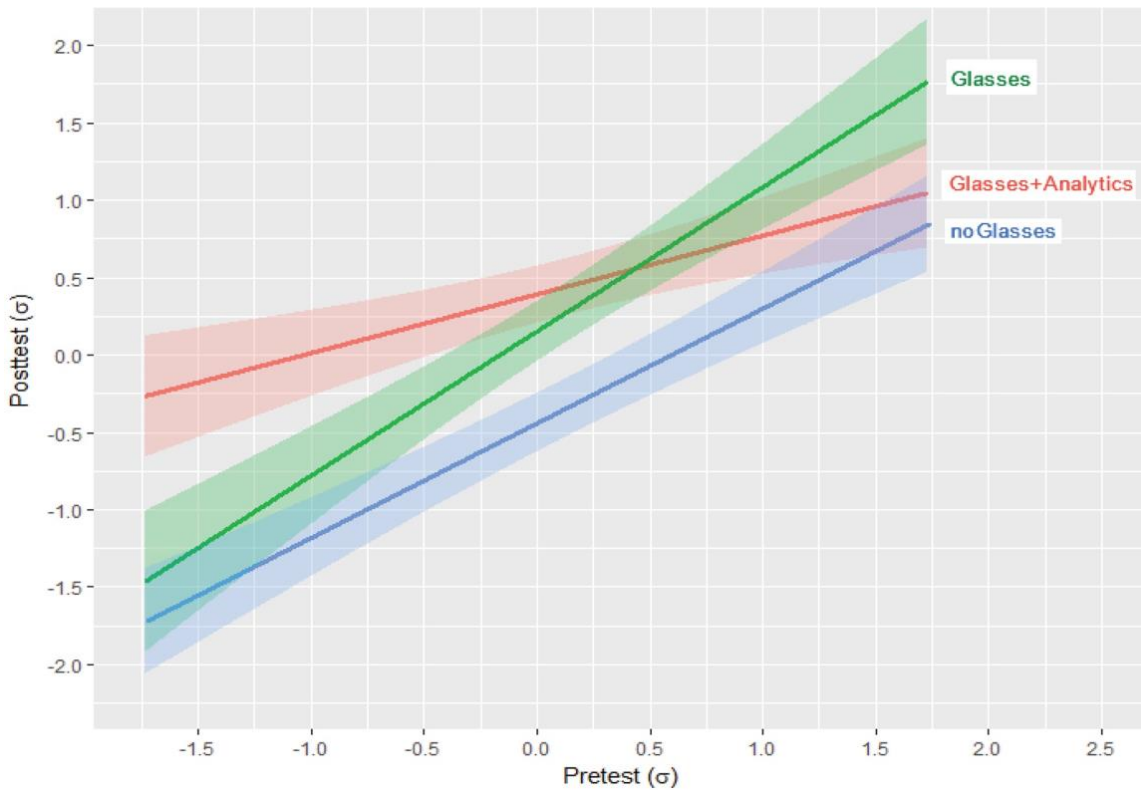

**Fig. 4.** Student posttest scores plotted by pretest scores, across 286 students. Lines indicate model mean values for each experimental condition; shaded regions indicate standard error; overlapping shaded regions indicate overlapping standard errors.

## DISCUSSION

Our case study illustrates the design and evaluation of a successful human–AI partnership in a challenging real-world context: K-12 education. While AI has shown great potential to enhance learning and teaching in K-12, the work of human teachers has been recognized as unlikely to be fully automated (*7, 24*). We concur: as in other care-based professions where relationship building is central, AI systems appear to have the greatest potential for positive impact where they are designed to *augment* and *complement* the abilities of human practitioners. In the Lumilo project, we took a participatory approach to research and design, deeply engaging teachers both in framing the challenges to be addressed through an improved partnership and in shaping how this partnership would function. Through iterative piloting in live K-12 classrooms, we observed the interplay of human and AI decision-making under real-world conditions, refining the design of Lumilo based on our observations to shape human–AI decision-making in more positive directions. Throughout this process, we worked to develop our understanding of what complementary strengths human teachers and AI tutors hold in our context, and how Lumilo's design might be optimized to effectively combine these strengths.

We view these components – taking a participatory approach to research and design, measuring and shaping human–AI decision-making in context, and working towards a theory of complementarity – as essential ingredients in the design of effective human–AI partnerships, both within and beyond the domain of education. Accordingly, to advance scientific and design progress in this area, we highlight three broad recommendations for future research. First, to support the development of human–AI partnerships that are aligned with real-world needs and work practices, it is critical to engage practitioners throughout the entire design and development lifecycle for a new technology. However, deeply involving practitioners becomes challenging when designing data-driven AI systems, given that practitioners may know very little about AI. The current case study illustrates the value of developing new design and prototyping methods. However, further research is urgently needed to develop new technical and design methods that can meaningfully engage non-technical stakeholders in understanding and working with AI as a design material, e.g., see (*12, 40*). Second, to support more reliable scientific and design insight into the behavior and dynamics of human–AI partnerships, future research should strive to study these systems in real-world contexts, with authentic tasks and relevant human experts. Although the use of artificial tasks and contexts may appear convenient, recent work has demonstrated that results from such studies often fail to translate to the real-world settings for which partnerships are intended, e.g., see (*4–5, 25*). Third, to support cumulative scientific progress and more systematic design exploration in this area, future research should work to develop and build upon both domain-general and domain-specific *theories of human–AI complementarity*. While we have begun to pursue this direction within the current case study, theory formation remains a critical open direction for the field, e.g., see (*5, 10, 17, 37*).

# Authors

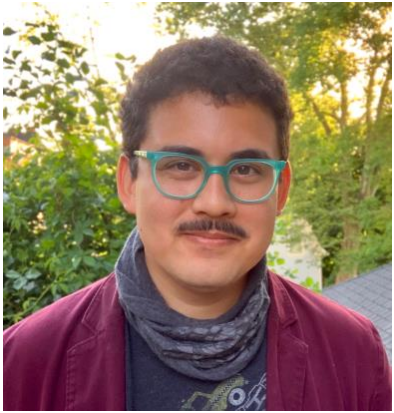

**Dr. Kenneth Holstein** (kjholste@cs.cmu.edu) is an Assistant Professor in Human-Computer Interaction (HCI) at Carnegie Mellon University, where he leads the Co-Augmentation, Learning, and AI (CoALA) Lab (https://www.thecoalalab.com/). His group studies how humans and AI systems can augment each other's abilities (co-augmentation) and learn from each other (co-learning) to support more effective and responsible human-AI collaborations. Throughout their work, they partner with stakeholders in real-world contexts, and create new methods to facilitate stakeholder involvement across the AI development lifecycle. Finally, they conduct field experiments to understand the impacts of these human–AI systems in deployment.

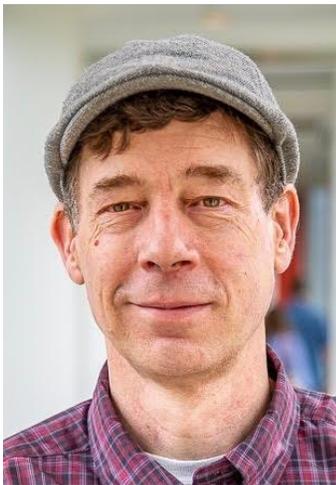

**Dr. Vincent Aleven** (aleven@cs.cmu.edu) is a Professor of Human-Computer Interaction (HCI) at Carnegie Mellon University. He has 25 years of experience in research and development of AI-based tutoring software grounded in cognitive theory and self-regulated learning theory. His current focus is on developing dynamic learning environments that combine strengths of human and AI instruction. His lab created the Mathtutor website (https://web.mathtutor.cmu.edu) as well as CTAT+Tutorshop, a key infrastructure for development of AI-based learning software, which has been used to create many dozens of tutoring systems that have been used in real educational settings. Aleven serves as co-editor-in-chief of the International Journal of Artificial Intelligence in Education (IJAIED).